\documentstyle[preprint,eqsecnum,aps,epsfig]{revtex}

\begin{document}
\setlength{\textwidth}{27pc}
\setlength{\textheight}{43pc}

\title{Many-polaron states in the Holstein-Hubbard model}
\maketitle

\begin{center}
 Laurent Proville 
and Serge Aubry${^*}$ \\
\small Groupe de Physique des Solides, UMR 7588-CNRS\\
 Universit\'es Paris 7 $\&$ Paris 6, Tour 23,
 2 pl. Jussieu 75251, Paris Cedex 05, France\\

${^*}$ Laboratoire L\'eon Brillouin (CEA-CNRS), CEA Saclay
        91191-Gif-sur-Yvette Cedex, France
\end{center}

\begin{abstract}
A variational approach is proposed to determine some properties of 
the adiabatic Holstein-Hubbard model which describes the interactions between a static atomic lattice and 
an assembly of fermionic charge carriers.
The sum of the electronic energy and 
the lattice elastic energy 
is proved to have  minima with a polaron structure in
a certain domain of the phase diagram.
Our analytical work consists in the expansion of these energy minima 
from  the zero  electronic transfer limit
which  remarkably holds for a finite amplitude of the onsite Hubbard repulsion
and for an unbounded lattice size.

\end{abstract}


\section{Introduction}

The Holstein model involves an electron-phonon coupling \cite{emin}
which yields a polaron
ground-state provided the amplitude of this coupling is large enough compared with the electronic transfer integral. 
This result was demonstrated by S. Aubry {\it et al.} in Ref. \cite{AAR92} by
assuming that the lattice modes have a negligible kinetic energy.  
In this adiabatic regime, the description of phonons is given by the static atomic displacements.
During the last decade, the so called small polaron of the adiabatic Holstein model
has been of a great interest to investigate some challenging
problems such as: charge density wave 
\cite{QCA92,RA95,lorenzo} and high critical temperature superconductivity \cite{PA98,PA99,PA00}. 
These studies were based on
the rigorous Aubry's proof \cite{AAR92} which has been improved 
by C. Baesens and R. S. MacKay in Ref. \cite{MB95}. 
In the anti-integrable (AI) limit where the transfer integral of the fermionic charges is zero
(see \cite{aubry-AI} for a revue about the anti-integrability), it is possible to
find the explicit forms of the Hamiltonian eigenstates and to expand them
numerically with respect to the charge transfer.

With no electron-electron interaction, excepted the Pauli principle, a many-electron
problem is usually reduced to find the eigenstates
of one particle Hamiltonian in which is introduced a chemical
potential to fix the charge density (see \cite{AAR92,lorenzo} for the case of the Holstein model). 
The ground state is thus given by the product of the  one particle eigenstates,  energies of which are smaller than the Fermi level.
In some cases, the interplay of the Coulombian interaction with a charge
screening may yield an onsite Hubbard repulsion which breaks this method. 
After extending the Holstein model with the
Hubbard interaction, it is yet possible to develop a variational approach where the energy functional $F_{ad}$ 
is issued from the bracketing of the adiabatic Holstein Hubbard Hamiltonian.
Minimizing this functional with respect to the displacement fields, for a  zero charge transfer t shows that some
energy minima are the products of onsite polaron, onsite  bipolaron 
and vacuum site. For the weak value of t,
the study of the first derivatives of $F_{ad}$ with respect to the atomic displacements
indicates at least one energy minimum with a displacement field which is similar to a AI polaron structure.
In the $L_{\infty}$ norm meaning, the distance between 
this minimum and the AI polaron state
is bounded by a charge transfer function which goes to zero continuously.
As a consequence, in the vicinity of  the AI limit, the displacement fields of the
many-polaron minima  are proved to evolve continuously with t which implies that the
$F_{ad}$ potential has some polaron-type minima in a finite domain of the phase diagram.

\section{Adiabatic Holstein Hubbard model}

The Holstein Hubbard Hamiltonian is written as follows:
\begin{equation}
{\cal H}=\sum_{i}\hbar
\omega _{0}(a_{i}^{\dag}a_{i})+gn_{i}(a_{i}^{\dag}+a_{i})
+{\upsilon }n_{i,\uparrow
}n_{i,\downarrow }-\nu n_{i} 
-T\sum_{ (i:j),\sigma }C_{i,\sigma }^{\dag}C_{j,\sigma } \label{hamilMB}
\end{equation}
where the atomic lattice is mapped on ${\mathcal Z}^d$, i.e., $i \in {\mathcal Z}^d$ with $d=\{1,2,3\}$,
the anihilation operators at site $i \in {\mathcal Z}^d$ for both phonons
and fermions are $a_{i}$ and $C_{i}$, respectively. 
The corresponding creation operators are written with a dag exponent. The 
$n_{i}$ operator is given by $\sum_{\sigma=(\uparrow,\downarrow)}
C_{i,\sigma }^{\dag}C_{i,\sigma }$ and the sum $\sum_{(i:j)}$ is performed over the $i$ neighboring site indexes. 
For simplicity, the phonon contribution is only one optical phonon branch with frequency $\omega_{0}$. 
The electron-phonon coupling amplitude is g, the onsite Hubbard repulsion is scaled by $\upsilon >0$ and
the  chemical potential is noted $\nu$.
The displacement and momentum operators at site $(i)$ are given by:

\begin{eqnarray}
&u_i&=\frac{\hbar\omega_0}{4g}(a_{i}^{\dag}+a_{i})  \label{position} \\
&p_{i}&=i\frac{2g}{\hbar\omega_0}(a_{i}^{\dag}-a_{i})  \label{momentum}
\end{eqnarray}
Substituting the phonon operators in Eq.(\ref{hamilMB}) 
and dividing this equation by the energy parameter $E_{0}=\frac{8g^{2}}{\hbar \omega _{0}}$ give:

\begin{equation}
H = \sum_i{\left(\frac{1}{2} u_i^2 + (\frac{u_i}{2}-\mu
) n_i +U
n_{i\uparrow}n_{i\downarrow} \right)} - t\sum_{(i:j),\sigma}
C_{i,\sigma}^{\dag}C_{j,\sigma} + \frac{\gamma}{2} \sum_{i}p_i^2
\label{Hh2}
\end{equation}
where $H$ is an adimensional Hamiltonian with the parameters:
\begin{equation}
U=\frac{\upsilon }{E_{0}}\qquad t=\frac{T}{E_{0}}\qquad \gamma =\frac{1}{4}(%
\frac{\hbar \omega _{0}}{2g})^{4}\;\;\;
\mu =\frac{\nu }{E_{0}} 
\label{param}
\end{equation}
In the adiabatic limit, the coefficient $\gamma$ is neglected which is
valid at large electron-phonon coupling, i.e., the amplitude
$g$ is large compared with the phonon zero point energy $\hbar \omega_0 /2$. 
We obtain an adiabatic Hamiltonian:

\begin{equation}
H_{ad}=\sum_{i}{\left( \frac{1}{2}u_{i}^{2}+(\frac{u_{i}}{2}-\mu
)n_{i}+Un_{i\uparrow }n_{i\downarrow }\right) }-{t}
\sum_{(i:j),\sigma }C_{i,\sigma }^{\dag}C_{j,\sigma }
\label{Had}
\end{equation}
The displacement operators $\{u_{i}\}$ are now  scalar variables that are
noted as a vector $\vec u \in {\cal S}(N)$ where ${\cal S}(N)={\cal R}^N$ is a real space, dimension
of which is equal to the unbounded number of lattice sites N. 
The adiabatic Hamiltonian is a sum of an
electronic Hamiltonian $H_{el}$ and the lattice elastic energy, i.e.,
$H_{ad}=\sum_{i}\frac{1}{2}u_{i}^{2}+H_{el}$ where
\begin{equation}
H_{el}=\sum_{i}(\frac{u_{i}}{2}-\mu )n_{i}+Un_{i\uparrow }n_{i\downarrow }-
t \sum_{(i:j),\sigma }C_{i,\sigma }^{\dag}C_{j,\sigma }  \label{Hel}
\end{equation}
The energy is now written in the variational form:

\begin{equation}
F({\vec u},\psi )=\sum_{i}\frac{1}{2}u_{i}^{2}+<\psi |H_{el}|\psi >\label{unon}
\end{equation}
where $|\psi >$  is a normalized electronic wave function
for N$_{el}$ fermion charges. It is
projected on the usual fermion basis, i.e., 
$|\psi >=\sum_{\nu} \psi_\nu |e_\nu>$ where 
$\nu$ is a N$_{el}$-multiple
site-spin index and 
$|e_\nu>=\Pi_{(i,\sigma_i)\in \nu} \
C^{\dag}_{i,\sigma_i} |\emptyset>$.
Differentiating  Eq.(\ref{unon}) with respect to  $u_{i}$ and $\psi_\nu$, the conditions for
the local extrema are:
\begin{eqnarray}
u_{i} &=&-\frac{<\psi |n_{i}|\psi >}{2}  \label{densmb} \\
H_{el}|\psi >&=& E |\psi >  \label{Hel1}\\
\nonumber
\end{eqnarray}
where $E$ is the Lagrange factor due to the $\psi$
normalization.
The Shroedinger equation Eq.(\ref{Hel1}) implies that $\psi $
is a $H_{el}$ 
eigenstate with the associated eigenvalue $E$ which is the electronic energy. 
Assuming that $\psi$ is a $H_{el}$ ground state
for a given $\vec u$, 
the total energy is now given by the
functional $F_{ad}({\vec u})=\sum_{i}\frac{1}{2}u_{i}^{2}+E(u_{i})$ which
depends only on the displacement variables. 
The $\psi$ wave function may be non-unique.

Working with the $L_\infty$ norm in the suitable space,
the $F$ functional and its derivatives with respect to $u_i$ and
$\psi_\nu$ are continue. The adiabatic potential $F_{ad}$ is also 
continue in the $\cal S$(N) space but its first derivatives are not necessary continue as it can be
shown in the  AI limit.

\section{The anti-integrable limit}

In the anti-integrable (AI) limit, 
the fermionic charge transfer is zero, and thus the
whole lattice sites are decoupled. 
The Hamiltonian $H_{el}$ is a sum of
onsite Hamiltonian $H_{i}$ and the  $H_{el}$ ground state $\psi $ is a product of 
onsite eigenstate $\pi_{i}$ with the energy $E_{i}(u_{i})$:

\begin{eqnarray}
H_{i} &=&(\frac{u_{i}}{2}-\mu)
n_{i}+U n_{i\uparrow }n_{i\downarrow } \label{Honsite} \\
H_{i}|\pi _{i}>&= &E_{i}(u_{i})|\pi _{i}> \label{Honsite2}\\ \nonumber
\end{eqnarray}
The constants $(\mu,U)$ fix which type of state is the $H_{i}$ ground state for a given displacement $u_i$.
Selecting a set of $N_{el}$ onsite states which have the lowest energy $E_i$  provides an electronic 
ground state $\psi$ for $H_{el}$.
The site $i$ is occupied either by 
a bipolaron, i.e., 2 electrons with opposite spin or 
by a polaron, i.e., only 1 electron with spin up or down or else the site i is not occupied.
Using Eq. \ref{densmb}, the optimum displacement field is such that $u_i=-1$ for a bipolaron,
or  $u_i=-1/2$ for a polaron or $u_i=0$ if the site i is unoccupied. 
In the case of a   bipolaron onsite ground state  with the optimum displacement 
$u_i=-1$, the $H_i$ eigenvalues are $(U-2\mu-1,-\mu-1/2,0)$.
For a polaron, the $H_i$  eigenvalues are   $(U-2\mu-1/2,-\mu-1/4,0)$
and for a vacuum site $(U-2\mu,-\mu,0)$.
In order to determine whether a polaron structure is
a local $F_{ad}$ minimum, it is sufficient to test the $H_i$ ground state 
for the different values of the onsite displacement $u_i \in \{-1,-1/2,0 \}$.
As example, for $\mu=-3/32$ and $U=3/8$,  
the $H_i$ ground state is a bipolaron for $u_i=-1$, 
a polaron for $u_i=-1/2$,  and it is a vacuum for $u_i=0$, so any displacement field consisting of an assembly of $u_i \in \{-1,-1/2,0 \}$
is a $F_{ad}$ minimum. On the opposite, if $\mu=-1/8$ and $U=1$, the configuration which
contains at least a  displacement $u_i = -1$ are not stable because the bipolaron  is no longer the $H_i$ ground state.

Let us note $\delta_i$ the  onsite spectrum gap between the ground state energy and the first
excited state energy. This gap is non-zero $\delta_i >0$ excepted for some specific values of $(\mu,U)$.
As a consequence  for nearly all $(\mu,U)$ constants, 
the onsite  ground state is not degenerate excepting the spin degeneracy which
occures for the polaron.

\section{Expansion of polaron structures}

As soon as the transfer integral $t$ is non zero, one may guess that
the AI polaron states
should still be $F_{ad}$ minima, at least for a certain range of parameter. 
Here we propose a proof that confirmes this guess.
Our demonstration is based on the $F_{ad}$  gradient study in the space of 
the displacement configurations ${\cal S}$(N). 
The $F_{ad}$ gradient is given by: 
\begin{equation}
\phi_{i}=u_{i}+\frac{<\psi |n_{i}|\psi >}{2} 
\end{equation}
Let us introduce the following operators:
\begin{eqnarray}
P_{1,i} &=&n_{i\uparrow } n_{i\downarrow }  \nonumber \\
P_{2,i}&= &n_{i\uparrow }(1-n_{i\downarrow })  \nonumber \\
P_{3,i}&= &n_{i\downarrow }(1-n_{i\uparrow })  \nonumber \\
P_{4,i}&= &1-n_{i\downarrow }-n_{i\uparrow }+n_{i\uparrow }n_{i\downarrow } 
\label{Projectors}
\end{eqnarray}
They verify $\sum_{\alpha } P_{\alpha ,i} \mid \psi >=\mid \psi >$ for all $\psi$ state.
We choose to note $P_{\alpha ,i} \mid \psi >=x_{\alpha,i}\mid \psi _{\alpha ,i}>$ where the state $\psi_{\alpha ,i}$ is
normalized and $x_{\alpha,i}$ is a real positive scalar. 
The gap $\delta_i$ is assumed to have a lower bound $\delta$, i.e., $\delta_i>\delta>0$ 
which is valid for nearly all $(\mu,U)$ constants.
Then only one $x_{\alpha ,i} $
is non-zero at the AI limit and it is equal to one, the corresponding index $\alpha$ is noted $g_i$.
If $g_i$ is equal to either 1 or 4 then we note $x_{g_i}=x_{\alpha=g_i,i}$. 
If $g_i$ is equal to either 2 or 3, the site i is occupied by a polaron with a spin up or down. In such a case, the 
onsite ground state is spin-degenerate and we note $x_{g_i}=\sqrt{ x_{2,i}^2+x_{3,i}^2}$.
As soon as the electronic transfer is non-zero, $x_{g_i}$
variates with t and it can be proved  that (see Lemma \ref{Lemme}):

\begin{equation}
(1-x_{g,i}^{2})^{\frac{1}{2}}<\frac{2 n_c n_{s}^{\frac{1}{2}}{t}}{\delta
-t n_c (n_{s}-2)}  \label{INEG1}
\end{equation}
where $n_c$ is the number of the nearest neighbors, 
$n_{s}$ is the number of distinct $H_{i}$ eigen values. 
The inequality Eq. (\ref{INEG1}) implies the continuity of $x_{g,i}$ with respect to $t$
in the vicinity of the AI limit.

The  potential $F_{ad}$ is now derived with respect to $u_i$:
\begin{equation}
\phi_{i}= \frac{\partial F_{ad}}{\partial u_i}=u_{i}+\sum_{\alpha ,\beta }x_{\alpha ,i}.{x}_{\beta ,i}.<\psi
_{_{^{\alpha ,i}}}|\frac{n_{i}}{2}|\psi _{_{^{\beta ,i}}}>  \label{phii}
\end{equation}
As $P_{\alpha , i} P_{\beta,i} = \delta_{\alpha,\beta} 
P_{\alpha, i}$ where $\delta_{\alpha,\beta} $ is the Kroeneker symbol and 
as $n_{i}$ commutes with the  $P_{\alpha,i}$ operators: 
\begin{equation}
\phi_{i}=u_{i}+\frac{1}{2}\sum_{\alpha }{x}_{\alpha ,i}^{2}<\psi_{\alpha
,i}|{n}_{i}|\psi_{\alpha ,i}>  \label{GRAD}
\end{equation}
Let note $n_{\alpha,i}=<\psi_{\alpha ,i}| {n}_{i} |\psi_{\alpha ,i}>$ 
and let write $u_{i}=u_{i}(0)+\rho _{i}$ where $u_{i}(0)$ is the onsite
displacement at $t=0$. The Eq.(\ref{GRAD}) is rewritten as follows
\begin{equation}
\phi_{i}=\rho_{i}-\frac{n_{g,i}}{2}(1-x_{g_i}^{2})+\frac{1}{2}
\sum_{\alpha \neq g_i} x_{\alpha,i}^{2} n_{\alpha,i}  \label{phii2}
\end{equation}
where we used Eq. (\ref{densmb}) 
in the AI limit to find $u_{i}(0)=-\frac{n_{g_i}}{2}$. 
The sum in the right hand side  of Eq. (\ref{phii2})  is performed 
over the indexes $\alpha \ne g_i$ for $g_i=\{1,4\}$ and over the indexes
$\alpha=\{1,4\}$ for $g_i=\{2,3\}$.

The scalar product  $\overrightarrow{\rho }.\overrightarrow{\phi }$ is now detailed:
\begin{equation}
\overrightarrow{\rho }.\overrightarrow{\phi }=\sum_i \rho _{i}^2+
\rho _{i}.(
-(1-x_{g_i}^{2})\frac{n_{g_i}}{2}+ \sum_{\alpha \neq g_i} x_{\alpha ,i}^{2} \frac{n_{\alpha,i}}{2} )  \label{INE22}
\end{equation}
Focussing on the terms of the right hand side sum, some simple arguments give the following inequalities:
\begin{eqnarray}
\rho _{i}.(1-x_{g_i}^{2})\frac{n_{g_i}}{2} &<& sup_i (1-x_{g_i}^{2})  |\rho _{i}| \\
\rho _{i}.  \sum_{\alpha \neq g_i} x_{\alpha ,i}^{2} \frac{n_{\alpha,i}}{2} &>& - sup_i (1-x_{g_i}^{2})  |\rho _{i}| \\
\nonumber
\end{eqnarray}
So each term of the sum in the Eq. (\ref{INE22}) verifies
\begin{equation}
\rho _{i}^2+\rho _{i}.(
-(1-x_{g_i}^{2})\frac{n_{g_i}}{2}+ \sum_{\alpha \neq g_i} x_{\alpha ,i}^{2} \frac{n_{\alpha,i}}{2} )> \rho _{i}^2- 2 |\rho _{i}|.
sup_i (1-x_{g_i}^{2})   \label{INE222}
\end{equation}
which is positive if $|\rho_i| > 2 sup_i (1-x_{g_i}^{2}) $ and with Eq. (\ref{INEG1}) it is equivalent to
$|\rho_i|   >  R_{AI}$ with writing:  
\begin{equation}
R_{AI}=\frac{8 n_c^{2}n_{s}{t}^{2}}{(\delta-t n_c(n_{s}-1))^{2}} \label{INE223}
\end{equation}
Defining the subset ${\cal B}(R) $  such as ${\vec u} \in {\cal B}(R)$ if $|u_i(0) - u_i| < R$ for all i index,
the boundary of ${\cal B}(R) $ is  noted ${B}(R) $. For any $\vec u \in B(R>R_{AI})$, the   product
$({\vec u}(0)-{\vec u}).{\vec \phi}({\vec u})$ is positive 
and thus there is at least one displacement configuration
${\vec u}_{min}(t) \in {\cal B}(R_{AI})$ which is a local  minimum of $F_{ad}$.
As a consequence,
The potential $F_{ad}$ has a minimum in the ${\vec u(0)}$
vicinity at most at a $R_{AI}$ distance in the $L_{\infty}$ norm meaning, i.e., 
$sup_i |u_{min,i}(t) - u_{i}(0) | < R_{AI}$.


\section{Conclusion}
The present proof holds for any displacement field
which is a minimum of the adiabatic potential $F_{ad}$ in the AI limit and such as the gap $\delta \ne 0$.
For nearly all
$(\mu,U)$ values, the displacements of the polaronic minima
variate continuously with respect to the fermion transfer t 
in the vicinity of the AI limit where t=0. So it is about the total energy of these minima
because of the $F_{ad}$ continuity with respect to the displacements.
Consequently, the adiabatic potential has some minima with a many-polaron structure in a finite region of the phase diagram.  
Nevertheless, the absolute minimum of $F_{ad}$, i.e., the ground state of Holstein-Hubbard model
can not yet be determined for all the parameters. To that purpose, 
a numerical investigation might be required but
no idea emerge to tackle the case of a non zero Hubbard coupling with many charge carriers, excepted to compute a
meanfield as proposed  in Ref.\cite{AAR92} or to study a model with a small electron number.
The latter possibility is presented in \cite{PA98,PA99} where
the phase diagram is calculated for 2 electrons. For a
two-dimensional atomic lattice, 
a critical point where coexist 3 different types of bipolaron was found far from any trivial limit. 
In this specific region, because of the bipolaron degeneracy, 
the bipolaron tunneling (or equivalently the inverse of the ground state effective mass)
is very 
sensitive to the quantum fluctuations which are yielded by a non-zero $\gamma$ (Eq. (\ref{param}))  Ref.\cite{PA00}.
Around the critical point, both
the bipolaron ground state mobility and its binding energy reach $100 K$
with realistic input parameters (Eq. (\ref{param})).
This result allowed some conjectures about the mechanism which yields
the high critical temperature superconductivity of cuprates.

The present study can be extended straighforwardly
for a non harmonic phonon potential as  for an atomic lattice
embedded in an external magnetic field. 
However an extension to a different electron-phonon coupling such as 
the SSH model \cite{SSH} might have non trivial anti-integrable limit
which makes our arguments much less efficient. 

\section{Lemma}
\label{Lemme}

We choose the $H_i$ ground state energy as the
energy reference. We shall assum the non-degeneracy of this ground state so
only one projector $P_{g_i}$ is such as
$H_{i}P_{g_i}|\psi >=0$ for all $\psi$. 
As $|\psi >=\sum_{\alpha }{P_{\alpha,i}}|\psi >$ we have
$|\psi >=\sum_{\alpha }{x_{\alpha,i }}|\psi _{\alpha,i}>$
where is introduced the normalized states $\psi_{\alpha ,i}$ and 
$x_{\alpha,i}|\psi _{\alpha ,i}>=P_{\alpha ,i}|\psi >$ which implies 
$\sum_{\alpha }{|x_{\alpha ,i}|}^{2}=1$.
It is possible to choose the $\psi_{\alpha ,i}$ such as 
the $x_{\alpha ,i}$ are real positive for all $\alpha$.
One notes $K_{i}=\sum_{\alpha } {P_{\alpha,i }}H_{el}{P_{\alpha,i}}$ and
$H_{{\bar \i}}=H_{el}-H_{i}+t \sum_{i:j,\sigma }C_{i,\sigma
}^{+}C_{j,\sigma }+C_{j,\sigma }^{+}C_{i,\sigma }$ 
where $(i:j)$ are the $i$ neighboring sites. 
As $P_{\alpha,i }[{C}_{i,\sigma}^{+}C_{j,\sigma }
+C_{j,\sigma }^{+}C_{i,\sigma }]P_{\alpha,i }=0$ and 
as $H_{{\bar \i}}$ and $H_{i}$ commute with the  projectors $P_{\alpha,i }$:

\begin{equation}
K_{i}=H_{{\bar \i}}+H_{i}  \label{HiH}
\end{equation}
Let $\Phi_{0}$ be the $K_{i}$ ground state with energy $E_0$. 
Using $P_{\alpha , i} P_{\beta,i} = \delta_{\alpha,\beta} 
P_{\alpha, i}$ where $\delta_{\alpha,\beta} $ is the Kroeneker symbol,
then $K_{i} P_{\alpha,i }|\Phi_{0}>=
P_{\alpha,i } K_{i} |\Phi_{0}>=E_{0} P_{\alpha,i }|\Phi_{0}>$. 

As $H_{{\bar \i}}$ and $H_i$ are decoupled, 
the state $\phi_0$ is a product of the $H_{\bar i}$ ground state 
and the $H_i$ ground state. 
It follows that 
$P_{g_i }|\Phi_{0}>=|\Phi_{0}>$ and $P_{\alpha\ne g_i }|\Phi_{0}>=0$
and it is now easy to establish that
\begin{eqnarray}
<\Phi_{0}|H_{\bar \i}|\Phi_{0}>&=&  E_{0} \label{aa0}\\
<\Phi_{0}|H_{el}|\Phi_{0}>&=& <\Phi_{0}|K_{i}|\Phi_{0}>=E_{0}  \label{edr1}\\
<\psi_{g_i}|K_{i}|\psi_{g_i} >& \ge& E_0 \label{edr2}\\
\nonumber
\end{eqnarray}

From the identity $P_{\alpha,i}^2=P_{\alpha,i}$, 
one deduces that for all $\psi$,
 $|\psi_{\alpha,i}>=P_{\alpha,i} |\psi>$, $P_{\alpha,i}
|\psi_{\alpha,i}>=|\psi_{\alpha,i}>$ and $<\psi_{\alpha,i}|K_i|\psi_{\alpha,i}>=<\psi_
{\alpha,i}|H_{el}|\psi_{\alpha,i}>$. If $\delta_i$ is the first excited state energy of  $H_i$:

\begin{eqnarray}
<\psi_{\alpha \ne g_i,i}|H_{i}|\psi_{\alpha \ne g_i,i}>&\ge & \delta_i  \label{lastEqm2} \\
<\psi_{\alpha \ne g_i,i}|K_{i}|\psi_{\alpha \ne g_i,i}> &=&<\psi_{\alpha \ne g_i,i
}|H_{{\bar \i}}|\psi_{\alpha \ne g_i,i}>+<\psi_{\alpha \ne g_i,i}|H_{i}|\psi_{\alpha \ne g_i,i}> \label{lastEqm1} \\
<\psi_{\alpha\ne g_i,i}|K_{i}|\psi_{\alpha \ne g_i,i}> &\geq &<\psi_{\alpha
\ne g_i,i}|H_{{\bar \i}}|\psi_{\alpha \ne g_i,i}>+\delta_i \geq  E_0 +\delta_i   \label{lastEq}  \\  \nonumber 
\end{eqnarray}
The previous results Eq. (\ref{lastEq}) is necessary valid if $\psi$ is the $H_{el}$ ground state:
\begin{equation}
<\psi_{\alpha \ne g_i,i}|K_{i}|\psi_{\alpha \ne g_i,i}>\geq {E}_{0}+\delta_i
\label{inegal2bis}
\end{equation}
Multiplying $H_{el}$ by the identity $(\sum_{\alpha }{P_{\alpha,i}}=Id)$
gives: 
\begin{equation}
H_{el}=K_{i}+\sum_{\alpha \neq \alpha^{\prime}} P_{\alpha,i }H_{el}{
P_{\alpha',i}}
\end{equation}
and bracketing by $\psi$:
\begin{equation}
<\psi|H_{el}|\psi>=\sum_\alpha{x^2_{\alpha,i}}
<\psi_{\alpha,i}|K_i|\psi_{\alpha,i}>
+\sum_{\alpha\neq\beta}x_{\alpha,i}{x_{\beta,i}}<\psi_{\alpha,i}
|H_{el}|\psi_{\beta,i}>
\label{popo}
\end{equation}
As  $n_{i,\sigma}$ and $n_{j,\sigma^{\prime}}$ commute  with each other for
all suffix $i,j$, $\sigma$ and $\sigma^{\prime}$, and as
$P_{\alpha,i}P_{\beta\neq \alpha,i}=0$:

\begin{equation}
\sum_{\alpha\neq\beta}x_{\alpha,i} {x_{\beta,i}}
<\psi_{\alpha,i}|H_{el}|\psi_{\beta,i}>=
-t\sum_{\alpha\neq\beta}x_{\alpha,i}
{x_{\beta,i}}<\psi_{\alpha,i}|\Delta|\psi_{\beta,i}>
\end{equation}
One deduces a simplification of the Eq. (\ref{popo}): 
\begin{equation}
<\psi|H_{el}|\psi>=\sum_\alpha{x^2_{\alpha,i}}<\psi_{\alpha,i}|K_i|
\psi_{\alpha,i}>
-t\sum_{\alpha\neq\beta}x_{\alpha,i}{x_{\beta,i}}
<\psi_{\alpha,i}|\Delta|\psi_{\beta,i}>
\end{equation}
and combining the inequalities (\ref{edr2},\ref{inegal2bis}):

\begin{equation}
<\psi |H_{el}|\psi >\geq {E}_{0}+\delta_i \sum_{\alpha \neq g_i}x_{\alpha,i
}^{2}-t\sum_{\alpha \neq \beta }x_{\alpha,i }{x}_{\beta,i }<\psi _{\alpha,i
}|\Delta |\psi _{\beta,i }>  
\end{equation}

As we choose to map the atomic lattice on ${\mathcal Z}^d$,
the number of first neighboring sites is 
$n_c=2 d$. With Eq. (\ref{edr1}), one now writes the set of following equations where $E=<\psi|H_{el}|\psi>$:
\begin{eqnarray}
<\Phi_{0}|H_{el}|\Phi_{0}> &\geq &E  \nonumber\\
E_{0}\geq E\ \rightarrow \ E_{0}\geq E &\geq &{E}_{0}+\delta_i \sum_{\alpha
\ne g_i}x_{\alpha,i }^{2}-t\sum_{\alpha \neq \beta }x_{\alpha,i }{x}_{\beta,i }<\psi
_{\alpha,i }|\Delta |\psi _{\beta,i }>  \nonumber\\
<\psi_{\alpha,i }|\Delta |\psi_{\beta,i }> &=&<\psi_{\alpha,i
}|\sum_{i:j,\sigma }C_{i,\sigma }^{+}C_{j,\sigma }+C_{j,\sigma
}^{+}C_{i,\sigma }|\psi _{\beta,i }>\leq 2n_{c} \nonumber\\
\delta_i \sum_{\alpha \ne g_i}x_{\alpha,i }^{2} &\leq & t \sum_{\alpha \neq \beta
}x_{\alpha,i }{x}_{\beta,i }<\psi _{\alpha,i }|\Delta |\psi _{\beta,i }>\nonumber\\
\delta_i (1-x_{g_i}^{2}) &\leq & 2n_{c}t\sum_{\alpha \neq \beta }x_{\alpha,i }{x}_{\beta,i } \nonumber\\
(1-x_{g_i}^{2})\delta_i &\leq & n_{c}t[(\sum_{\alpha }x_{\alpha,i})^{2}-\sum_{\alpha }x_{\alpha,i }^{2}]  \nonumber\\
(1-x_{g_i}^{2})\delta_i &\leq & n_{c}t[(\sum_{\alpha }x_{\alpha,i })^{2}-1] 
\nonumber \\
(1-x_{g_i}^{2})\delta_i & \leq &n_{c}t[(\sum_{\alpha \ne g}x_{\alpha,i } +x_{g_i})^{2}-1] \label{inegal4} \\
\nonumber
\end{eqnarray}
The Cauchy-Schwartz inequality applied to the
sum $\sum_{\alpha \ne g}x_{\alpha,i }$ gives: 
\begin{eqnarray}
\sum_{\alpha \ne g} x_{\alpha,i } &\leq & \sqrt{\sum_{\alpha\ne g_i} 1}
\sqrt{\sum_{\alpha \ne g_i}x_{\alpha,i }^{2}}  \nonumber \\
\sum_{\alpha \ne g_i}x_{\alpha,i } &\leq &\sqrt{n_{s}}\sqrt{1-x_{g_i}^{2}}  \nonumber
\end{eqnarray}
where 
$n_{s}=3$ is the maximum number of distinct $H_{i}$ eigenvalues.
With Eq. (\ref{inegal4}), it follows that
\begin{equation}
(1-x_{g_i}^{2})\delta_i \leq n_{c}t[(n_{s}^{\frac{1}{2}}(1-x_{g_i}^{2})^{\frac{1}{2}}+x_{g_i})^{2}-1] 
\end{equation}
and it is now easy to obtain:

\begin{equation}
(1-x_{g_i}^2)^\frac{1}{2}<\frac{2 n_c n_s^\frac{1}{2} t} 
{\delta_i- n_c t (n_s-1)} \label{noDEGx}
\end{equation}

The latest result holds for the case of a non-degenerate $H_i$ ground state.
Regarding the case of a polaron at site i, the atomic orbital is occupied by 1 electron with either spin up or spin down.
The onsite ground state is spin-degenerate but
the same arguments as for the non-degenerate case
can be used to establish the following inequality:
\begin{equation}
(1-x_{2,i}^2-x_{3,i}^2)^\frac{1}{2}<\frac{2 n_c (n_s-1)^\frac{1}{2} t} 
{\delta_i- n_c t (n_s-2)} 
\end{equation}
Let write $x_{g_i}^2=x_{2,i}^2+x_{3,i}^2$ such as this inequality is now written:
\begin{equation}
(1-x_{g,i}^2)^\frac{1}{2}<\frac{2 n_c (n_s-1)^\frac{1}{2} t} 
{\delta_i- n_c t (n_s-2)}  \label{DEGx}
\end{equation}
If we now assum that there is a lower bound $\delta$ for the onsite gap $\delta_i>\delta>0$, then
the combination of the Eqs. (\ref{noDEGx}, \ref{DEGx}) gives for all site i :
\begin{equation}
(1-x_{g_i}^2)^\frac{1}{2}<\frac{2 n_c n_s^\frac{1}{2} t} 
{\delta- n_c t (n_s-2)}  \label{DEGx2}
\end{equation}

\end{document}